\documentclass[reprint,amsmath,amssymb,aps,twocolumn,superscriptaddress]{revtex4-2}
\usepackage{graphicx}
\usepackage{dcolumn}
\usepackage{bm}
\usepackage{float}
\usepackage{amssymb}
\usepackage{color}
\usepackage{multirow}
\usepackage{stmaryrd}
\usepackage[colorlinks,linkcolor=blue,urlcolor=blue,citecolor=blue]{hyperref}

\begin{document}


\title{Flat optical conductivity in topological kagome magnet TbMn$_6$Sn$_6$}

\author{R. S. Li}
\thanks{These authors contributed equally to this work.}
\affiliation{International Center for Quantum Materials, School of Physics, Peking University, Beijing 100871, China}
\author{Tan Zhang}
\thanks{These authors contributed equally to this work.}
\affiliation{Beijing National Laboratory for Condensed Matter Physics, Institute of Physics, Chinese Academy of Sciences, Beijing 100190, China}
\author{Wenlong Ma}
\affiliation{International Center for Quantum Materials, School of Physics, Peking University, Beijing 100871, China}
\author{S. X. Xu}
\affiliation{International Center for Quantum Materials, School of Physics, Peking University, Beijing 100871, China}
\author{Q. Wu}
\affiliation{International Center for Quantum Materials, School of Physics, Peking University, Beijing 100871, China}
\author{L. Yue}
\affiliation{International Center for Quantum Materials, School of Physics, Peking University, Beijing 100871, China}
\author{S. J. Zhang}
\affiliation{International Center for Quantum Materials, School of Physics, Peking University, Beijing 100871, China}
\author{Q. M. Liu}
\affiliation{International Center for Quantum Materials, School of Physics, Peking University, Beijing 100871, China}
\author{Z. X. Wang}
\affiliation{International Center for Quantum Materials, School of Physics, Peking University, Beijing 100871, China}
\author{T. C. Hu}
\affiliation{International Center for Quantum Materials, School of Physics, Peking University, Beijing 100871, China}
\author{X. Y. Zhou}
\affiliation{International Center for Quantum Materials, School of Physics, Peking University, Beijing 100871, China}
\author{D. Wu}
\affiliation{Beijing Academy of Quantum Information Sciences, Beijing 100913, China}
\author{T. Dong}
\affiliation{International Center for Quantum Materials, School of Physics, Peking University, Beijing 100871, China}
\author{Shuang Jia}
\affiliation{International Center for Quantum Materials, School of Physics, Peking University, Beijing 100871, China}
\affiliation{Collaborative Innovation Center of Quantum Matter, Beijing 100871, China}
\affiliation{CAS Center for Excellence in Topological Quantum Computation, University of Chinese Academy of Sciences, Beijing 100190, China}
\author{Hongming Weng}
\email{hmweng@iphy.ac.cn}
\affiliation{Beijing National Laboratory for Condensed Matter Physics, Institute of Physics, Chinese Academy of Sciences, Beijing 100190, China}
\affiliation{School of Physical Sciences, University of Chinese Academy of Sciences, Beijing 100049, China}
\affiliation{Songshan Lake Materials Laboratory, Dongguan, Guangdong 523808, China}
\affiliation{CAS Center for Excellence in Topological Quantum Computation, Beijing 100190, China}
\author{N. L. Wang}
\email{nlwang@pku.edu.cn}
\affiliation{International Center for Quantum Materials, School of Physics, Peking University, Beijing 100871, China}
\affiliation{Beijing Academy of Quantum Information Sciences, Beijing 100913, China}
\affiliation{Collaborative Innovation Center of Quantum Matter, Beijing 100871, China}

\date{\today}

\begin{abstract}
Kagome magnet TbMn$_6$Sn$_6$ is a new type of topological material that is known to support exotic quantum magnetic states. Experimental work has identified that TbMn$_6$Sn$_6$ hosts Dirac electronic states that could lead to topological and Chern quantum phases, but the optical response of the Dirac fermions of TbMn$_6$Sn$_6$ and its properties remain to be explored. Here, we perform optical spectroscopy measurement combined with first-principles calculations on single-crystal sample of TbMn$_6$Sn$_6$ to investigate the associated exotic phenomena.
TbMn$_6$Sn$_6$ exhibits frequency-independent optical conductivity spectra in a broad range from 1800 to 3000 cm$^{-1}$ (220-370 meV) in experiments. The theoretical band structures and optical conductivity spectra are calculated with several shifted Fermi energy to compare with the experiment. The theoretical spectra with 0.56 eV shift for Fermi energy are well consistent with our experimental results. Besides, the massive quasi-two-dimensional (quasi-2D) Dirac bands, which have linear band dispersion in $k_x$-$k_y$ plane and no band dispersion along the $k_z$ direction, exist close to the shifted Fermi energy. According to tight-binding model analysis, the quasi-2D Dirac bands give rise to a flat optical conductivity, while its value is smaller than, about one tenth of, that from the calculations and experiments. It indicates that the other trivial bands also contribute to the flat optical conductivity.

\end{abstract}

\pacs{Valid PACS appear here}
\maketitle


\section{\label{sec:level1}INTRODUCTION}
Kagome material has attracted immense interest due to its special crystal structure, which serves as an important platform to investigate the interplay among magnetism, electron correlation effects, and topological order~\cite{keimer2017physics, haldane1988model, xu2015intrinsic,yin2018giant, yin2019negative}. The kagome lattice has a corner-sharing triangles network and it is predicted to have dispersionless bands and band singularities (van Hove singularities). 
Many intriguing novel phenomena and properties, such as anomalous Hall effect~\cite{haldane2004berry, nakatsuji2015large, wang2016anomalous, liu2018giant}, topological Dirac states~\cite{ye2018massive}, quantum spin liquid states~\cite{zhou2017quantum, balents2010spin}, charge density wave, and superconductivity~\cite{ortiz2019new} have been discovered in those materials. 
In a kagome lattice, combining ferromagnetic (FM) ordering with strong spin-orbit coupling (SOC) can break time-reversal symmetry~\cite{xu2015intrinsic}, which yields nontrivial topological states. Intrinsic anomalous Hall effect can arise from the integration of Berry curvature over the Brillouin zone (BZ)~\cite{haldane2004berry}. Additionally, kagome magnets with broken time-reversal symmetry are supposed to be an attractive candidate for the realization of the spinless Haldane model~\cite{haldane1988model, xu2015intrinsic}. The discovery of superconductivity and charge density wave in the kagome material CsV$_3$Sb$_5$ promises to be important progress in condensed-matter physics~\cite{ortiz2019new,ortiz2020cs}. It is confirmed that the kagome lattice is a fertile platform to explore novel states and phase behaviors.

Recently, the transition-metal-based kagome magnet RMn$_6$Sn$_6$ (where R is a rare earth element) family has been studied extensively in the context of magnetism-induced various topological states and phase behaviors ~\cite{yin2020quantum, ma2020rare, ma2021anomalous, li2021dirac, xu2022topological, li2022manipulation, riberolles2022low, zhang2022exchange, lee2022interplay, jones2022origin, mielke2022low}. This class of materials is predicted to support the nontrivial topological states owing to a pristine Mn-base kagome layer with weak interlayer coupling. While Mn atoms form magnetic kagome lattice, R atoms also contribute to magnetization, which plays a significant role in topological properties of RMn$_6$Sn$_6$~\cite{venturini1991magnetic, ma2020rare}. As all of the atom moments contribute to net $c$-axis ferromagnetic ordering without any magnetic impurities, TbMn$_6$Sn$_6$ is an ideal system to search for the Chern gapped topological fermions. As a recent article reported, TbMn$_6$Sn$_6$ can host Landau quantization on the application of a 9 T magnetic field, which is a moderate magnetic field for kagome material to reach quantum limit~\cite{yin2020quantum}.

TbMn$_6$Sn$_6$ has a layered hexagonal crystal structure, with a manganese kagome layer stacked along the $c$-axis. 
Out-of-plane ferrimagnetic order can arise from anti-parallel coupled magnetic coupling between Tb and Mn moment below Curie temperature \emph{T}$_c$ = 423 K~\cite{MALAMAN1999519, el1991magnetic}.
Due to net magnetization and SOC, Dirac bands open a gap at the crossing points that gives rise to the correlated massive Dirac fermions. Combining scanning tunneling microscope (STM) and angle-resolved photoemission spectroscopy (ARPES) measurements, TbMn$_6$Sn$_6$ indeed host linear dispersions near the zone corners with a Chern gap above the Fermi level~\cite{yin2020quantum}. The energy from the top of the lower Dirac branch to the Fermi level is about 130 meV, while the Chern gap $\Delta$ is about 34 meV according to Ref.~\onlinecite{yin2020quantum}. Massive Dirac fermions generate large Berry curvature that induces intrinsic anomalous Hall effect. In addition, quantum oscillation data indicated that this Fermi surface in the bulk crystal is two-dimensional like~\cite{yin2020quantum,xu2022topological}. 
However, because of the complex Mn-3$d$ electron correlation effect, density function theory calculations could not give accurate energy band structures, where the calculated location of  quasi-two-dimensional (quasi-2D) Dirac bands is inconsistent with the STM/ARPES results~\cite{lee2022interplay, jones2022origin,yin2020quantum}. 
In the work of Yin \emph{et al.}~\cite{yin2020quantum}, the calculated Fermi level has an unspecified shift of 0.56 eV to match the experimental results, which is not a common case. Considering the effect of Mn-3$d$ electron correlation by DFT+U method, the quasi-2D Dirac cone at 0.7 eV above the Fermi level shifts down to 0.3 eV~\cite{lee2022interplay}, but it's still far away from the Fermi level to have a dominant contribution to the electrical response. 
Using density functional theory plus dynamical mean-field theory (DFT + DMFT) in YMn$_6$Sn$_6$~\cite{li2021dirac}, the calculated quasi-2D Dirac cone still occurs at about 300 meV above the Fermi energy. 
Furthermore, the Berry curvature of all the occupied states should contribute anomalous Hall effect. It is not reasonable to attribute all the topological features to the single Dirac cone at the $K$ point, especially when this Dirac cone is far from the Fermi level based on the raw calculation results~\cite{lee2022interplay}. In addition to these experimental and theoretical works, we would like to investigate the optical response of TbMn$_6$Sn$_6$, which involves both occupied and unoccupied states and is expected to give out more details in electronic structures. The real position of Dirac point needs to be acquired by direct experiment probes, which is important to understand the origin of its topological nature. Besides, while most bands near the Fermi level are highly dispersive, TbMn$_6$Sn$_6$ exhibit quasi-2D transport behaviors~\cite{xu2022topological}, which should also be further studied. Therefore, it is highly interesting to investigate the unusual electronic properties of TbMn$_6$Sn$_6$, which can elucidate the quantum nature of this kagome magnet.

In this work, we systematically investigate the underlying physics of kagome lattice TbMn$_6$Sn$_6$ by combining optical spectroscopy and first-principles calculations. Optical spectroscopy is a valuable and important method that probes charge dynamics and the electronic band structures of materials. Owing to the deeper penetration of light in the infrared range, optical spectroscopy reflects the bulk properties of a material, while STM and ARPES are sensitive to the surface. 
Additionally, the Dirac points and most dispersionless bands of TbMn$_6$Sn$_6$ are above the Fermi level and couldn't be detected by ARPES. Compared with the ARPES measurements, the advantage of optical probe is that it can measure the inter-band transition from the occupied state to the unoccupied state. 
As is known, when a system consists of Dirac fermions, the real part of the optical conductivity $(\sigma=\sigma_1+i\sigma_2)$ is supposed to follow a power law, i.e. $\sigma_1\sim \omega^{d-2}$ (where $d$ is dimension of the system)~\cite{hosur2012charge, bacsi2013low, tabert2016optical}. The power law is a compelling tool to investigate the properties of Dirac fermions and has indeed been confirmed in various systems. For 2D Dirac semimetal Fe$_3$Sn$_2$~\cite{biswas2020spin} and 2D Dirac nodal line semimetal ZrSiS~\cite{schilling2017flat}, the optical conductivity is independent with frequency. On the other hand, 3D Dirac semimetal ZrTe$_5$ ~\cite{chen2015optical}, NbAs$_2$~\cite{shao2019optical}, and 3D Weyl semimetal Co$_3$Si$_2$S$_2$~\cite{yang2020magnetization} display linear optical conductivity.

\section{\label{sec:level2}EXPERIMENTAL RESULTS}

\begin{figure}
\includegraphics[clip, width=0.5\textwidth]{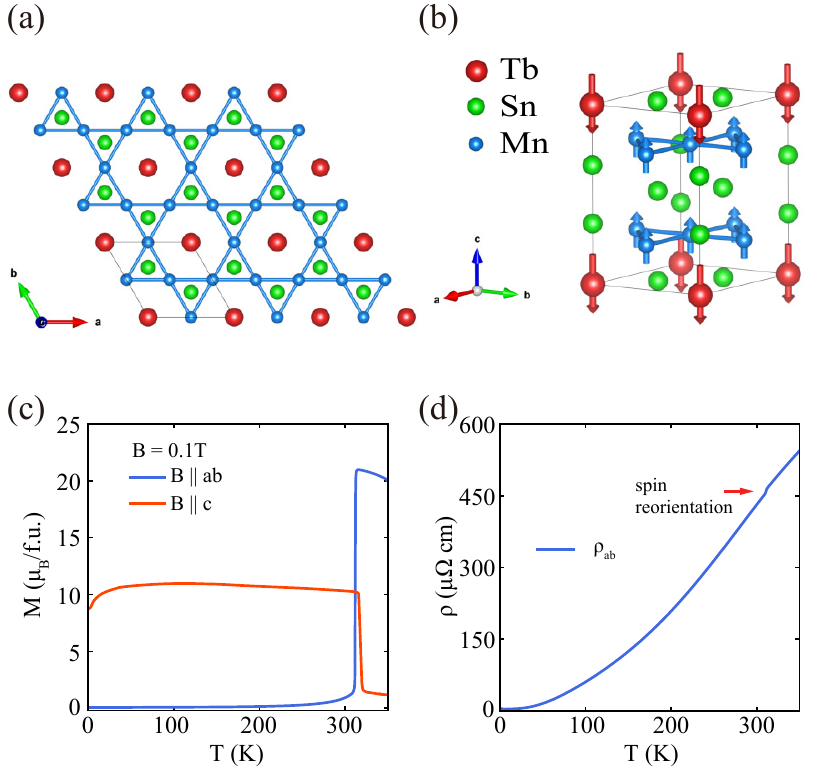}\\[1pt] 
\caption{Crystal structure, magnetic susceptibility, and resistivity in TbMn$_6$Sn$_6$. (a) Crystal of TbMn$_6$Sn$_6$ within a unit cell is shown by the gray solid lines. (b) Side view of the structure with the orientation of the Tb and Mn magnetic moments below 310 K. (c) The temperature-dependence magnetic susceptibility with a zero-field-cooling mode at  $B = 0.1$ T for $B \parallel ab$ and $B \parallel c$. (d) The dc-resistivity of TbMn$_6$Sn$_6$ as a function of temperature. The red arrow indicates the spin-reorientation transition at 310 K in the resistivity measurement.
\label{fig:1}}
\end{figure}

\begin{figure*}[htbp]
\includegraphics[clip, width=1\textwidth]{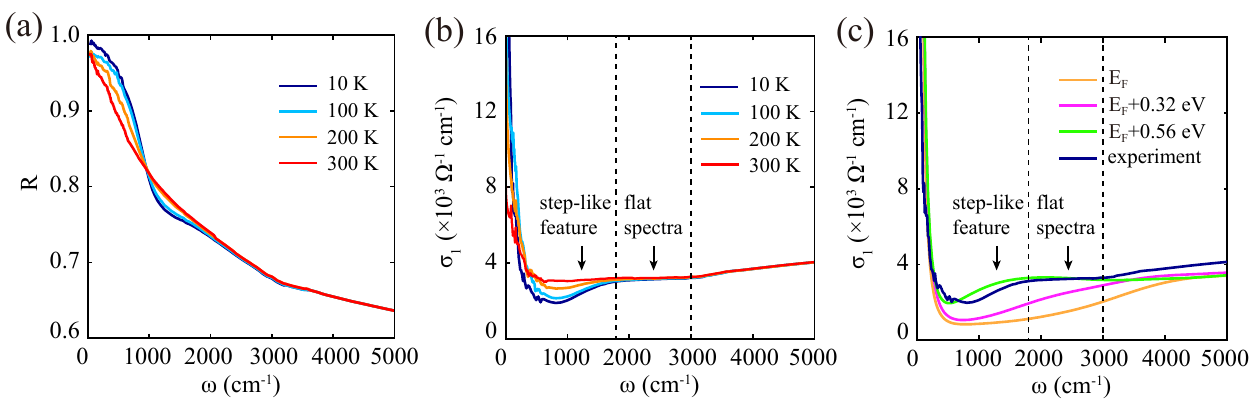}\\[1pt] 
\caption{(Color online) (a) Temperature-dependent optical reflectivity R($\omega$). (b) The experimental optical conductivity $\sigma_1$($\omega$) at different temperature.  (c) The theoretical optical conductivity $\sigma_1$($\omega$) at 0 K with shifts of Fermi energy and the experimental optical conductivity at 10 K. The black arrows indicate the step-like feature and flat optical conductivity, respectively.
\label{fig:optic}}
\end{figure*}

TbMn$_6$Sn$_6$ has a layered structure with space group $P6/Pmmn$ and hexagonal lattice constants of $a=b=5.5$ $\mathrm{\AA}$, $c=9$ $\mathrm{\AA}$. It is formed by a manganese kagome layer, with tin and terbium sequentially distributed in alternating layers stacked along the c-axis, as illustrated in Figs.~\ref{fig:1}(a) and (b). 
TbMn$_6$Sn$_6$ orders in a collinear in-plane ferrimagnetic spin structure below Curie temperature (\emph{T}$_c$ = 423 K)~\cite{el1991magnetic}. With the decrease in temperature, it manifests a spin reorientation at about 315 K, where the $ab$-plane aligned moments are flopped to be parallel to the $c$-axis~\cite{MALAMAN1999519, el1991magnetic, jones2022origin}, as reflected in the magnetic susceptibility anisotropy in Fig.~\ref{fig:1}(c).  The temperature-dependent magnetization experiments were performed in field-cooled conditions under a magnetic field of 0.1 T applied out-of-plane and in-plane, respectively. Both susceptibilities measured along the ab plane and $c$-axis show a prominent change around 315 K, featuring the spin reorientation~\cite{ jones2022origin}. Below this temperature, all of the Mn moments are aligned on the $c$-axis with Tb moments anti-aligned along the $c$-axis. 
Figure. ~\ref{fig:1} (d) shows the in-plane dc-resistivity as a function of temperature $T$, which was measured using a standard four-probe method in a Quantum Design physical property measurement system. In the entire measuring temperature range, the resistivity shows metallic behavior.  Notably, a slight anomaly in resistivity at 310 K arises from the spin reorientation.

Figure. ~\ref{fig:optic} (a) displays the temperature-dependent in-plane reflectivity up to 5000 cm$^{-1}$. Above 1800 cm$^{-1}$, the change of temperature has only a minor influence on the reflectivity spectra. The low-energy reflectivity reaches almost unity. Additionally, when the temperature decreases, the low-energy reflectivity increases. Both features show that the sample has a highly metallic nature, in agreement with the resistivity measurement. The most significant feature in spectra R($\omega$) is the substantial suppression in the mid-infrared region. Upon cooling, there appears a dip gradually emerges between 800 and 1500 cm$^{-1}$, which may indicate that this material opens a gap. 

The real part of the optical conductivity $\sigma_1$($\omega$) is derived from R($\omega$) through the Kramers-Kronig relation, as displayed in Fig.~\ref{fig:optic}(b). The Hagen-Rubens relation was used for the low-frequency extrapolation, and the X-ray atomic scattering functions were used for the high-frequency extrapolation. 
Let us first discuss the real part of the optical conductivity at 10 K. The conductivity consists of a sharp Drude peak at zero-frequency. Then a frequency-independent conductivity is followed by a step-like absorption feature, where the step-like feature ranges from 900 to 1800 cm$^{-1}$ as shown in Fig. ~\ref{fig:optic}(b) (we shall explain later). The striking feature of the spectra is that the flat frequency-independent region in the range from 1800 to 3000 cm$^{-1}$. Naturally, such a typical signature is supposed to be the contribution of quasi-2D Dirac fermion as predicted by the power laws~\cite{bacsi2013low, biswas2020spin, schilling2017flat}. Since the Dirac band loses its linearity and other parabolic bands start to have a dominant contribution to the conductivity, $\sigma_1(\omega)$ increases with frequency above 3000 cm$^{-1}$.

We now focus on the temperature dependence of the optical conductivity. 
The Drude peak narrows continuously with decreasing temperature, implying suppressed quasiparticle scattering. The optical conductivity above 1800 cm$^{-1}$ depends weakly on temperatures, where the plateau remains nearly constant across all measured temperatures. This behavior is consistent with temperature independence of Dirac energy bands derived from the $d_{xy}$ and $d_{x^2}$ orbitals of Mn atom. Because most of the Dirac energy bands arise from the spin-up orbitals of Mn atom, temperature change has a weaker effect on the magnetic moment of Mn atom. Notably, the step-like feature structure gets broad at high temperatures, which is similar to the behavior in Fe$_3$Sn$_2$~\cite{ biswas2020spin}.

\begin{figure}[H]
\includegraphics[clip, width=0.5\textwidth]{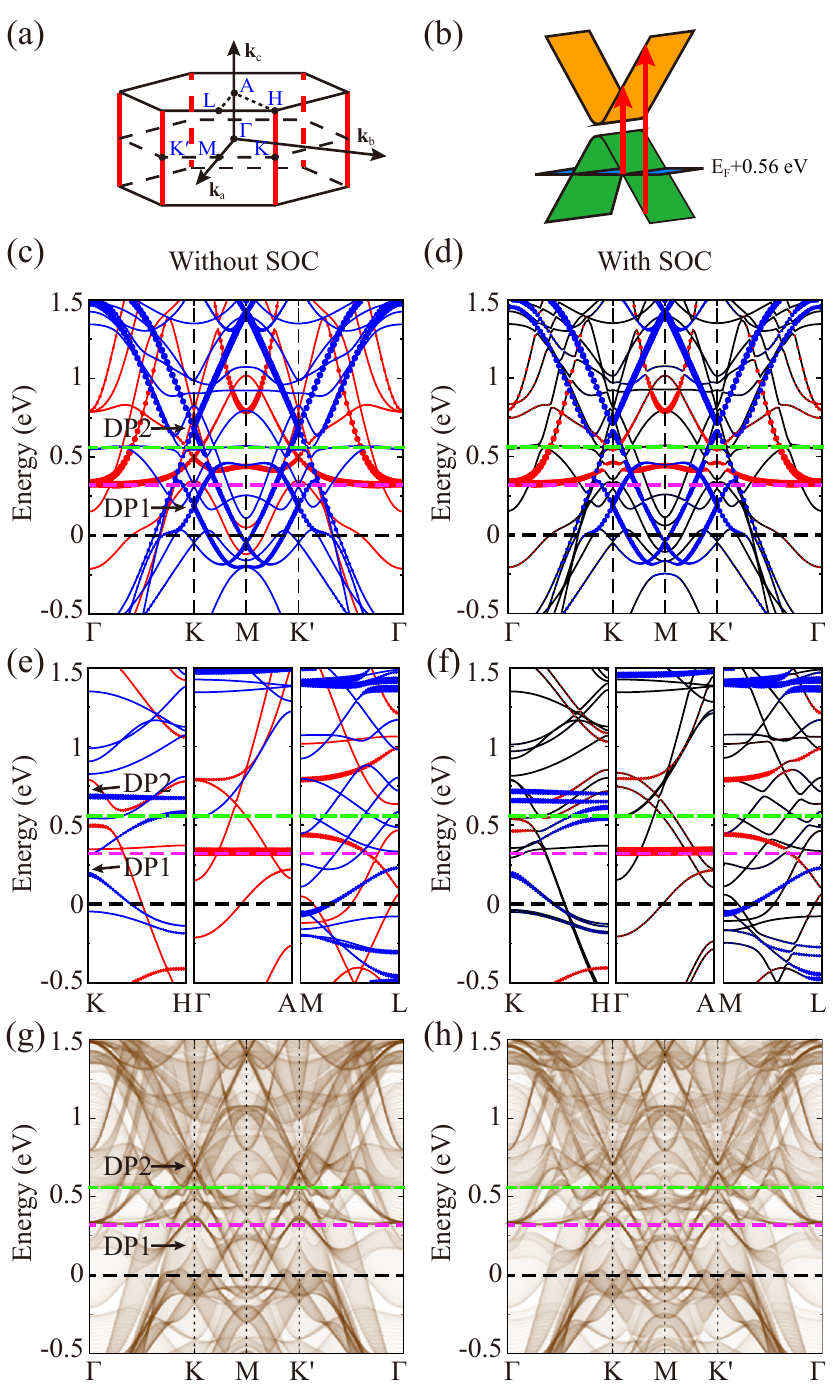}\\[1pt]
\caption{(Color online) (a) The bulk BZ of TbMn$_6$Sn$_6$. The red solid (dash) lines indicate the position of the nodal line. (b) The schematic diagram of the quasi-2D Dirac bands near the Fermi energy. The red arrows indicate the inter-band transitions. The band structures of TbMn$_6$Sn$_6$ in the $k_z = 0$ plane (c) without and (d) with SOC. The band structures along the $k_z$ path (e) without and (f) with SOC. The red (blue) bands indicate the spin-up (spin-down) bands. The red (blue) circles indicate the projection weight of spin-up (spin-down) Mn $d_{xy}$ and $d_{x^2}$ orbitals.  The overlapped band structures on 50 $k_z$ planes from $k_z=0$ to $0.5k_c$ (g) without and (h) with SOC. The deeper brown color indicates that more bands are overlapped. Two Dirac cones (DP1 and DP2) at the $K$ point are marked by arrows. The magenta and green dashed lines mark the 0.32 and 0.56 eV shift of Fermi energy.
\label{fig:bands}}
\end{figure}

To better understand the experimental results of TbMn$_6$Sn$_6$, the first-principles calculations are performed to simulate the band structures and optical conductivity. The TbMn$_6$Sn$_6$ hosts the ferrimagnetic state at 0 K, where the effective magnetic moments on Tb and Mn atoms are along the $z$-axis but in the opposite direction, as shown in Fig.~\ref{fig:1} (b). 
The magnetization on Tb atom is contributed by the 4$f$ orbitals which are away from the Fermi energy. To simplify the calculation, we treat the Tb-4$f$ electrons as core electrons and do not consider them in the DFT calculation by using the pseudopotential which does not contain 4$f$ electrons. The detailed calculation parameters are shown in Appendix \ref{ap1}.

The band structures in the $k_z=0$ plane without SOC are shown in Fig.~\ref{fig:bands}(c). There are two Dirac points (DP1 and DP2) at the energy of 0.187 and 0.682 eV located at $K$ and $K^{\prime}$ points. They have linear band dispersion in the $k_z=0$ plane, which consists of Mn spin-up $d_{xy}$ and $d_{x^2}$ orbitals. 
The band structures along the $k_z$ direction are shown in  Fig.~\ref{fig:bands}(e). It shows that the DP1 has a band dispersion along the $k_z$ direction, while the DP2 has no band dispersion along the $k_z$ direction. The DP2 makes up nodal lines with quasi-2D Dirac bands along the $k_z$ direction and goes through the whole BZ. The position of nodal lines is indicated as six red solid (dash) lines in Fig.~\ref{fig:bands}(a), which are related by $C_6$ rotation symmetry.
The band structures on 50 $k_z$ planes from $k_z=0$ to $0.5k_c$ are overlapped in Fig.~\ref{fig:bands}(g). The quasi-2D Dirac bands shown as deep brown color can be found at $K$ and $K^{\prime}$ points.
On considering SOC, the DP1 opens a small band gap (14 meV), while the DP2  opens a larger band gap (55 meV) at $K$ and $K^{\prime}$ points as shown in Fig.~\ref{fig:bands}(d). The Dirac points also open gaps along the $k_z$ direction (Fig.~\ref{fig:bands}(f)), but the band dispersion hardly change. The overlapped band structures with SOC are shown in Fig.~\ref{fig:bands}(h). The quasi-2D Dirac bands shown as deep brown color open a gap of 55 meV. To verify our calculation results, the bands in the $k_z$ = 0.3$k_c$ and 0.5$k_c$ planes are calculated as shown in Appendix \ref{ap2}.

Our calculations and the other similar calculations in Ref.~\onlinecite{lee2022interplay} show that the DFT band structures of TbMn$_6$Sn$_6$ are inconsistent with the STM/ARPES measurements in Ref.~\onlinecite{yin2020quantum}. The major discrepancy is the location of quasi-2D Dirac cone (DP2) in Fig.~\ref{fig:bands}(c). In DFT calculations it is 0.682 eV above the Fermi level, while it is about 0.125 eV according to Ref.~\onlinecite{yin2020quantum} through STM/ARPES measurements. One natural argument is the insufficient treatment of strong correlation effects of Mn 3$d$ electrons in DFT calculations since the Dirac cone is composed of Mn 3$d$ orbitals. However, our DFT+U calculations with various U and J parameters show this Dirac cone remains about 0.3 eV higher than Fermi level, which is consistent with Ref.~\onlinecite{lee2022interplay}. On considering these, we take the DFT calculation for the discussion and based on it the optical properties have been calculated with the up shift of Fermi level at the values of 0, 0.32 and 0.56 eV, respectively, as shown in Fig.~\ref{fig:optic}(c).

The theoretical optical conductivity at 0 K is calculated based on the band structures. The detailed calculation formulas are introduced in Appendix \ref{ap1}. The plasma frequency is calculated as $\hbar\omega_{p}$ = 2.95 eV. When the Fermi energy is shifted by 0.32 and 0.56 eV, the calculated plasma frequency changes to 3.94 and 4.09 eV accordingly. The inverse lifetime $\hbar\gamma$ is parameterized as 0.006 eV for all calculations. 
The theoretical optical conductivity spectra are shown in Fig.~\ref{fig:optic}(c). The spectra contributed by intra-band transitions are mainly in the region from 0 to 1000 cm$^{-1}$. When $\omega \geqslant 1000$ cm$^{-1}$, the spectra are dominated by the inter-band transitions. With the rising of Fermi energy, the spectra from inter-band transition gradually change from a linear dispersion with frequency to a step. When the Fermi energy is shifted by 0.56 eV, the theoretical spectra show a frequency-independent plateau existing from 1800 to 3000 cm$^{-1}$, which are well consistent with the experimental spectra as shown in Fig.~\ref{fig:optic}(c). The comparison between experimental and theoretical spectra indicates that the consistency is gradually improved when Fermi level is shifted up from 0 to 0.56 eV.

We think that this flat spectrum could be contributed by the inter-band transitions between quasi-2D Dirac bands, displayed in Fig.~\ref{fig:bands}(b). 
According to previous research, this optical conductivity can be calculated as an analytic formula by tight-binding model ~\cite{shao2019optical, PhysRevB.104.L201115,schilling2017flat, carbotte2016optical}. The formula is 
\begin{align}
\sigma_1=\frac{e^2k_0}{16h} \times \Theta(\hbar\omega-\text{max}\left\{ \Delta,2E \right\}),
\label{tbmodel}
\end{align}
where $k_0$ is the total length of quasi-2D Dirac bands in momentum space, $\Theta$ is the step function, $\Delta$ is zero (the gap of Dirac cone) and $E$ is the energy difference between the Fermi energy and the Dirac point (the midpoint of gap) for gapless (gapped) Dirac cone.
For a gapless Dirac cone, the frequency-independent optical conductivity exists from $\hbar\omega =0$ when the Dirac point locates at the Fermi energy. When the Dirac point is away from the Fermi energy, the optical conductivity starts to increase arises above $\hbar\omega \geqslant 2E$, because the inter-band transitions are forbidden for low frequency ($\hbar \omega< 2E$) due to Pauli blocking.  Such a gradual increase in conductivity is called a step-like feature.
For a gapped Dirac cone, the step-like optical conductivity arises when $\hbar\omega \geqslant \Delta$ if the Fermi energy is in the gap ($\Delta>2E$), and it arises when $\hbar\omega \geqslant 2E$ if the Fermi energy is away from the gap ($\Delta \leqslant 2E$).

In our calculations for TbMn$_6$Sn$_6$, the gap of quasi-2D Dirac band is $\Delta=55$ meV, while the midpoint of the gap is $E=124$ meV beyond the shifted Fermi energy. It infers the start point of optical conductivity step is at $2E=248$ meV by the formula Eq.~\ref{tbmodel}, which is consistent with that of 223 meV (1800 cm$^{-1}$) in our calculation. 
Also, the magnitude of optical conductivity $\sigma_1$ is calculated  by Eq.~\ref{tbmodel}. We estimated $\sigma_1=0.34 \times 10^3~\Omega^{-1}$cm$^{-1}$, as we calculated the length of nodal line $k_0$ = 1.4~\AA$^{-1}$ in the whole BZ.
It is about one tenth of the calculated value $\sigma=3.20 \times10^3~\Omega^{-1}$cm$^{-1}$ with contributions from all bands. We think the reason may be that Eq.~\ref{tbmodel} only estimates the conductivity plateau caused by Dirac cone bands, and the other overlapped bands make up a uniform band background in Figs.~\ref{fig:bands} (g) and (h) also contribute to constant optical spectra.

It should be noted that the calculated conductivity is closer to the experimental values when Fermi energy is shifted upward by 0.56 eV. Besides, the energy of quasi-2D Dirac cone after such a shift of Fermi energy is more in line with the previous experimental results, including STM, ARPES, and electrical transport measurement~\cite{yin2020quantum, xu2022topological}. As we pointed out above, the improved treatment of strong correlation effects of Mn 3$d$ electrons in DFT+U calculations only partially corrects the location of this Dirac cone. Therefore, the location of quasi-2D Dirac cone and its role in these physical phenomena is not conclusive within our optical conductivity analysis.

\section{\label{sec:level4}CONCLUSION}

In summary, we have carried out comprehensive optical spectroscopy and theoretical calculations to investigate the electron properties of the TbMn$_6$Sn$_6$ single crystal. The experimental optical conductivity spectra exhibit a frequency-independent plateau in a range from 1800 to 3000 cm$^{-1}$ (220-370 meV). This feature can be well described by our first-principles calculations when the Fermi level is shifted up by 0.56 eV as compared with those results from the shift value of 0 and 0.32 eV. The massive quasi-2D Dirac bands are found to exist close to the shifted Fermi energy as pointed out by former experiment. However, there are many other topologically trivial bands coexisting around the same energy region. Both the quasi-2D Dirac bands and these trivial bands can contribute to the observed and calculated optical conductivity. Within our calculations and analysis, the contribution from quasi-2D Dirac bands is very small, about one tenth of the total. Therefore, the plateau feature in optical conductivity is not necessarily reflecting the Dirac cone like linear dispersion. Careful analysis from both experiment and calculation is demanded.

\begin{acknowledgments}
This work was supported by National Natural Science Foundation of China (Grant Nos. 11888101, 11925408, 11921004 and 12188101), the National Key Research and Development Program of China (Grant Nos. 2022YFA1403900, 2017YFA0302904, 2018YFA0305700 and 2022YFA1403800), the Chinese Academy of Sciences (Grant No. XDB33000000) and the Informatization Plan of Chinese Academy of Sciences (Grant No. CAS-WX2021SF-0102).
\end{acknowledgments}

\appendix
\section{EXPERIMENT METHOD}\label{ap0}
TbMn$_6$Sn$_6$ single crystals were grown using the self-tin-flux method~\cite{canfield1992growth,clatterbuck1999magnetic}. Terbium pieces, manganese pieces and tin drops were mixed in a molar ratio of 1:6:20, put into an alumina crucible and then sealed in an evacuated quartz ampoule. The ampule was quickly heated to 1000 $^{\circ}$C and kept for a few hours, then slowly cooled down to 600 $^{\circ}$C over several days, and centrifuged at this temperature to separate the crystals from the Sn flux. Several millimeter-sized crystals with metallic luster were obtained in the crucible. The crystal has a flake-like shape with a shining a-b plane as the cleavage plane, consistent with its quasi-2D structure.

The in-plane reflectivity R($\omega$) was measured by the Fourier transform infrared spectrometer Bruker 80 V in the frequency range from 50 to 32000 cm$^{-1}$. At low frequencies, an in situ gold overcoating technique was used to get the reflection. Instead of gold, aluminum served as a reference above 8000 cm$^{-1}$.
 
\section{CALCULATION METHOD}\label{ap1}
The First-principles calculations were performed to simulate the electronic structure of TbMn$_6$Sn$_6$ by using the Vienna ab initio Simulation Package (VASP)~\cite{PhysRevB.47.558,PhysRevB.49.14251,KRESSE199615}. The Perdew-Burke-Ernzerhof (PBE) exchange-correlation potential with the generalized gradient approximation (GGA) was used~\cite{PhysRevLett.77.3865}. The projector augmented plane-wave pseudopotential was chosen in the calculations ~\cite{PhysRevB.50.17953}. The cut-off energy was 500 eV and the reciprocal space was sampled by 13$\times$13$\times$7 $\Gamma$-centered mesh for self-consistent calculation. 

The optical conductive properties of the material were calculated based on the longitudinal expression of the dielectric matrix~\cite{PhysRevB.73.045112,PhysRevB.79.125117}. The $k$-point sampling mesh is 21$\times$21$\times$11 mesh. 
The real part of optical conductivity is calculated by the formula: 
\begin{align}
\sigma_1(\omega)=\frac{\omega}{4\pi}\operatorname{Im}\epsilon(\omega) \label{sigma1},
\end{align}
where dielectric function consists of inter-band and intra-band contributions,
\begin{align}
\operatorname{Im}\epsilon(\omega)=\operatorname{Im}\epsilon_{\text{inter}}(\omega) + \operatorname{Im}\epsilon_{\text{intra}}(\omega) \label{epsilon}.
\end{align}
The imaginary part of inter-band dielectric function are calculated by the formula:
\begin{equation}
\begin{aligned}
\operatorname{Im}\epsilon_{\text{inter}}(\omega)=
& \frac{8\pi^{2} e^{2}}{V} \lim _{|\mathbf{q}| \rightarrow 0} \frac{1}{|\mathbf{q}|^{2}} \sum_{\mathbf{k}, v, c}\left|\left\langle u_{c, \mathbf{k}+\mathbf{q}} \mid u_{v, \mathbf{k}}\right\rangle\right|^{2} \\
& \times \delta\left(\epsilon_{c, \mathbf{k}+\mathbf{q}}-\epsilon_{v, \mathbf{k}}-\hbar \omega\right) \label{inter}.
\end{aligned}
\end{equation}
The intra-band contributions are calculated by the Drude model:
\begin{align}
\operatorname{Im}\epsilon_{\text{intra}}(\omega)=\frac{\gamma\omega_{p}^2}{\omega(\omega^2+\gamma^2)} \label{intra}.
\end{align}

\section{THE BAND STRUCTURES IN THE OTHER $k_z$ PLANES}\label{ap2}

\begin{figure*}
\includegraphics[width=2\columnwidth]{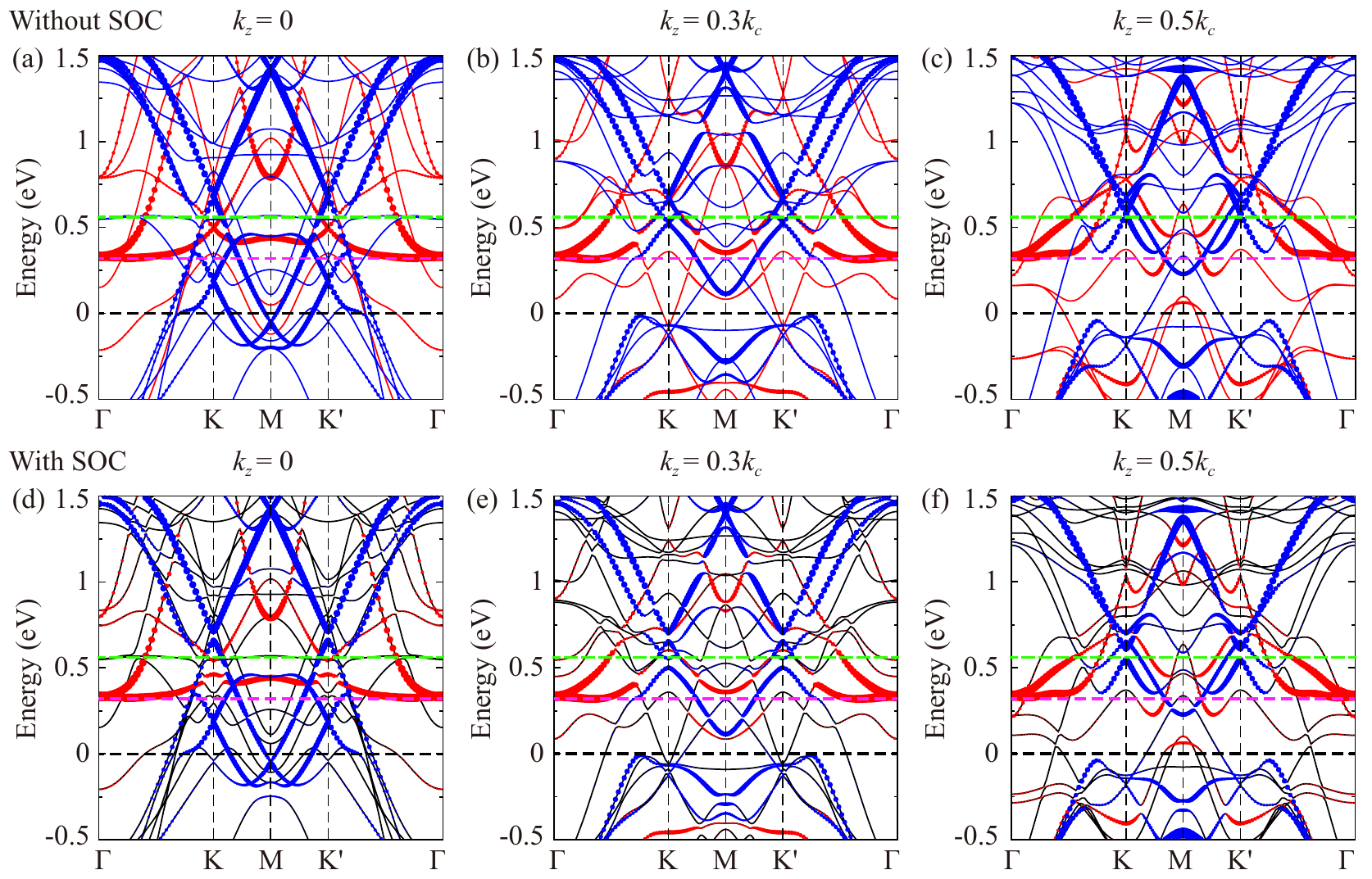}
\caption{(Color online) The band structures of TbMn$_6$Sn$_6$ in the $k_z = 0$ plane (a) without and (d) with SOC. The bands in $k_z = 0.3k_c$ plane (b) without and (e) with SOC. The bands in $k_z = 0.5k_c$ plane (c) without and (f) with SOC. The red (blue) circles indicate the projection weight of spin-up (spin-down) Mn $d_{xy}$ and $d_{x^2}$ orbitals. The magenta and green dashed lines mark the 0.32 and 0.56 eV shift of Fermi energy
\label{fig:bandkz}}
\end{figure*}

The band structures of TbMn$_6$Sn$_6$ in the $k_z$ = 0, 0.3$k_c$ and 0.5$k_c$ planes with and without SOC are shown in the Fig.~\ref{fig:bandkz}.

\bibliography{Ref}

\end{document}